\documentclass[useAMS,usenatbib]{mn2e}
\usepackage{captcont}
\usepackage{longtable}
\usepackage{graphicx}
\usepackage{natbib}
\usepackage{lscape}
\usepackage{rotating}
\usepackage{hyperref}
\usepackage{amssymb}
\usepackage{color}
\title[\textit{FAR ULTRAVIOLET SPECTROSCOPIC EXPLORER OBSERVATION} OF O~{\small VI} ABSORPTION IN THE MILKY WAY]{The distribution and kinematics of interstellar O~{\small VI} in the Milky Way}
\author[R. Sarma et al.]{R. Sarma$^{1}$\thanks{e-mail: sharma.rathin@gmail.com}, {A. Pathak$^{2}$\thanks{e-mail: amitpah@gmail.com}}, Jayant Murthy$^{3}$ and J. K. Sarma$^{2}$  \\
$^{1}$\textit{Department of Physics, Hojai College, Hojai, 782435, India}\\
$^{2}$\textit{Department of Physics, Tezpur University, 784028, India} \\
$^{3}$\textit{Indian Institute of Astrophysics, Koramangala, Bangalore 560 034, India}}
\begin{document}
\date{}
\pagerange{\pageref{firstpage}--\pageref{lastpage}}
\maketitle
\label{firstpage}
\begin{abstract}
We present the results of a survey of interstellar O~{\small VI} absorption in the Milky Way (MW) towards {69} stars in the Large Magellanic Cloud (LMC) obtained with the \textit{Far Ultraviolet Spectroscopic Explorer (FUSE)}. The integrated MW O~{\small VI} column densities log N(O~{\small VI}) are in the range from 13.68 to 14.73 with a mean of {14.26$_{-0.09}^{+0.07}$} atoms cm$^{-2}$. The O~{\small VI} exponential scale height is found to be {2.28$\pm$1.06} kpc. The O~{\small VI} column density correlates with the Doppler parameter $\it b$. The O~{\small VI} velocity dispersion ranges from {14.0 to 91.6 with an average value of 62.7 km s$^{-1}$}. These high values of velocity dispersion reveal the effect of turbulence, multiple velocity components and collision on broad O~{\small VI} profiles. There is a significant variation of O~{\small VI} column density on all scales studied {(0.0025$\degr$ - 6.35$\degr$)}. The smallest scale for which O~{\small VI} column density variations has been found is {$\bigtriangleup \theta$ $\sim$~9$\arcsec$}. Comparison of the O~{\small VI} velocity profiles with Fe~{\small II} indicates the presence of intermediate velocity cloud (IVC) and/or high velocity cloud (HVC) components in the O~{\small VI} absorption. 
\end{abstract}
\begin{keywords}
 ISM:atoms--ISM:galaxies:Milky Way
\end{keywords}
\section{Introduction}

{Highly ionized gas spanning the temperature range {from} 10$^{5}$ to 10$^{7}$ K is an important constituent of the interstellar medium (ISM) of galaxies. Species in the hot phase of the ISM (Si~{\small IV}, C~{\small IV}, N~{\small V}, O~{\small VI}, etc.) show strong transitions in the ultraviolet (UV). Important lines amongst these are the O$^{+5}$ (O~{\small VI}) absorptions at 1032 \AA\ and 1038 \AA. O~{\small VI} is unlikely to arise from photoioization as 113.9 eV is required to remove one more electron from O~{\small V} but is rather produced in the ISM through collisional ionization at temperatures of about 3$\times$10$^{5}$~K \citep{Indebetouw04, cox 2005}. Such temperatures represent the interface between the warm (T $\sim 10^{4}$ K) and the hot (T $>$ $10^{6}$ K) phases of the interstellar gas and a detailed analysis of the O~{\small VI} absorption aids in understanding the total abundance and the processes leading to its formation in the ISM \citep{Wakker12}.}

The first observations of O~{\small VI} absorption were made with the the {\it Copernicus} satellite \citep{Jenkins 1978a, Jenkins 1978b} followed by observations made with the {\it Hopkins Ultraviolet Telescope (HUT)} \citep{Dixon et al. 1996}. The  {\it Far Ultraviolet Spectroscopic Explorer (FUSE)} \citep{Moos et al. 2000, Sahnow et al. 2000} launched in 1999 June contributed significantly to the observation of O~{\small VI} absorption with much higher resolution and signal to noise. {\it FUSE} had a far-ultraviolet (FUV) wavelength coverage between  905 -- 1187 \AA\ with a resolution of $\sim$ 15,000 -- 20,000 near the O~{\small VI} line absorption wavelengths. The study of O~{\small VI} absorption in the ISM of galaxies with {\it FUSE} has given us information about the formation and distribution of O~{\small VI} in the MW and the Magellanic Clouds that adds to our knowledge of varying ISM conditions in environments of different metallicities (\citet{Howk02a, Savage00, Wakker03, Oegerle05, Savage06, Welsh08, Pathak et al.11}). Apart from absorption studies, {\it FUSE} has also been used to observe the O~{\small VI} spectra in emission from the diffuse ISM in the MW \citep{Shelton01, Dixon06, Dixon08}, from superbubbles (SBs) and in the LMC \citep{Sankrit07}. O~{\small VI} absorption studies at low redshifts trace the warm-hot intergalactic medium (WHIM) and are important contributor to models of the cosmological problem of missing baryons \citep{Tepper-Garcia11}.

\section{O~{\small VI} in the Milky Way}

Although the first observations of hot interstellar gas of the MW were made with the {\it Copernicus} satellite \citep{York, Jenkins 1978a}, it was not until the launch of the {\it FUSE} satellite that large scale observations of O~{\small VI} absorption in the MW began. Strong O~{\small VI} absorption was detected along 10 out of 11 lines of sight by Savage et al.\citet{2000} with log N(O~{\small VI}) ranging from 13.80 to 14.64 atoms cm$^{-2}$ while Howk et al.\citet{2002b} found a mean column density of 14.52$^{+0.10}_{-0.14}$ atoms cm$^{-2}$ in the direction of the LMC.
Savage et al.\citet{2005} report {\it FUSE} observations of O~{\small VI} absorption towards 100 extragalactic lines of sight finding that the average log N(O~{\small VI}) is 14.36 atoms cm$^{-2}$. Bowen et al.\citet{2008} studied absorption lines of O~{\small VI} towards 148 early-type stars situated at distances $\ge$1 kpc and found an average O~{\small VI} midplane density of 1.3$\times$10$^{-8}$cm$^{-3}$.

Pathak et al.\citet{2011} surveyed the  O~{\small VI} absorption in the LMC towards 70 lines of sight using {\it FUSE} and noted the presence of a MW component in each of these lines of sight. We have analysed all these sightlines and report our findings here.

\section{Observations and data reduction}

A detailed description of the {\it FUSE} mission and instruments has been given by Moos et al.\citet{2000} and Sahnow et al.\citet{2000}. {\it FUSE} consists of two channels (SiC and LiF) optimized for short and long wavelength observations. These channels are further divided into eight segments; the SiC 1A, SiC 2A, SiC 1B, SiC 2B covering the wavelength range 905 - 1105 \AA~and LiF 1A, LiF 2A, LiF 1B and LiF 2B covering the wavelength range 1000 - 1187 \AA.~{\it FUSE} observes through three apertures- LWRS, MDRS and HIRS with aperture size 30$^{\arcsec}\times 30^{\arcsec}$, 4$^{\arcsec}\times 20^{\arcsec}$ and 1.25$^{\arcsec}\times 20^{\arcsec}$ respectively. Here we use data from the LiF 1A segment since its sensitivity near 1032 \AA~is almost double to that of the other segments.

We obtained the calibrated data for the 69 sightlines analysed  by Pathak et al.\citet{2011} (see Table 1) from the Multimission Archive at STScI (MAST). These were processed by the latest {\it FUSE} data reduction pipeline CALFUSE V3.2 \citep{Dixon07}. Out of the 69 sightlines, O {\small VI} absorption has already been reported for 1 sight line by \citet{Friedman00}; for 11 sightlines by \citet{Howk02b}; 3 by \citet{Danforth06a}; and 1 by \citet{Lehner07}. Although the native {\it FUSE} resolution is 20 km~s$^{-1}$ in this region, we have uniformly downgraded all the spectra to a resolution of 35 km~s$^{-1}$ to have a higher signal-to-noise, irrespective of the quality of the original spectrum.

\subsection{Continuum and Contamination} 

Our first task was to define a stellar continuum near the O~{\small VI} profile for which we used Legendre polynomials as described by Sembach \& Savage\citet{1992} and Howk et al.\citet{2002a}. For most of the sightlines a low order polynomial ($\leq$5) is enough to model the continuum, except for a few sightlines (e.g., Sk-67D05, Sk-67D168, Sk-70D115 etc.) that show complex behaviour. These sightlines show a sudden hike or trough near the O~{\small VI} absorption. Polynomials of higher order ($>$5) were required to fit such continuum. To measure the error in the continuum fitting we have tested two continua other than the assumed continuum, one just above and the other just below the real continuum giving 1$\sigma$ significance level.

\begin{figure*}
\includegraphics[width=\textwidth]{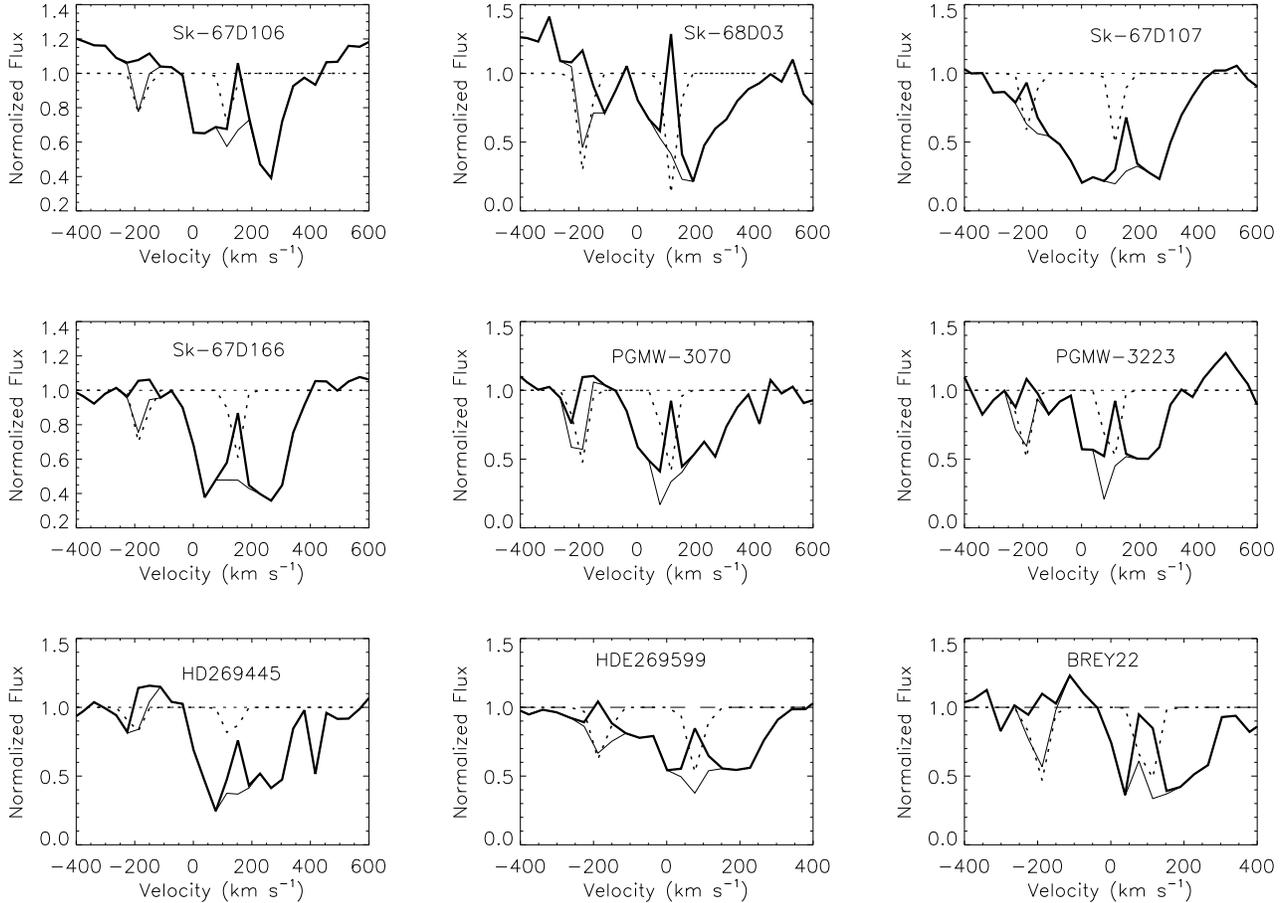}
\caption{O~{\small VI} absorption profiles for few lines of sight in the MW. Absorption line profiles of O~{\small VI} before (gray line) and after (black line) removal of H$_{2}$ contamination are shown. Model for H$_{2}$ absorption (dotted line) for these lines of sight are also shown.} 
\label{Fig1}
\end{figure*}

Another source of error in the O~{\small VI} column density measurements is the contamination from Galactic and LMC molecular hydrogen. We modeled the H$_{2}$ absorption lines that arise due to 6 -- 0 P(3) and 6 -- 0 R(4) transitions following Howk et al.\citet{2002a} (Table 3). {In this model we used the intensity of existing H$_{2}$ absorption lines that are not blended with other lines and use the ratio of the oscillator strengths.} The model estimated the strength of the 1031.20 \AA~and 1032.35 \AA~H$_{2}$ line. These were subtracted from the O~{\small VI} profiles that appear in the same wavelength region. This is not a major problem for the LMC absorption lines which are shifted to a higher velocity (\citet{Howk02a, Pathak et al.11}) but may affect the Galactic O~{\small VI} significantly. We have plotted a representative sample of normalized spectra of the O~{\small VI} absorption at 1032 \AA~in Figure \ref{Fig1}.  HD lines were modeled but we find that these are extremely weak and their contribution is within the error bars of O~{\small VI}.

{Stellar wind variability and mass loss may add significant error in the O~{\small VI} column density measurements. Interstellar O~{\small VI} identification and stellar continuum placement is complicated because of stellar wind variability. Following
\citet{Zsargo2003} we have selected by eye those the objects for which we had confidence in the continuum placement, line identification, and blending decontamination. \citet{Bowen et al.} studied the absorption spectrum of HD 178487 in detail and estimated the uncertainty due to stellar mass loss to be about 0.1 dex.}

\subsection{Apparent Optical Depth Measurements}

We have used the apparent optical depth (AOD) technique \citep{Savage91, Sembach92, Howk02a} to measure the equivalent width and column density of the O~{\small VI} absorption lines. This technique uses an apparent optical depth ($\tau_a$) in terms of velocity defined as

\begin{equation}
\tau_a(v)= ln[I_{o}(v)/I_{obs}(v)],
\end{equation}
where $I_o$ is the continuum intensity and $I_{obs}$ the intensity of the absorption line in terms of velocity. This method is well suited for observations where the instrument is able to completely resolve the absorption line.  

The apparent column density ($N_a(v)$ in units of atoms cm$^{-2}$ (km~s$^{-1}$)$^{-1}$) is calculated from the apparent optical depth using the expression
 
\begin{equation}
N_a(v) = \frac{m_e c \tau_a(v)}{\pi e^2 f \lambda} = 3.768 \times 10^{14} \frac{\tau_a(v)}{f \lambda},
\end{equation}
where $m_e$ is the mass of the electron, $c$ is the speed of light, $e$ is the electronic charge, $\lambda$ is the line wavelength (in \AA) and {\it f} is the oscillator strength of the atomic species. For O~{\small VI}, {\it f} value of 0.1325 has been taken from \citet{Yan98}. 
The O~{\small VI} absorption is a doublet with absorptions at 1032 and 1038 \AA. While the 1032 \AA\ O~{\small VI} profile is broad and is completely resolved by {\it FUSE}, the 1037.6 \AA\ is difficult to separate from the CII* absorption line. Because of this we have only used the {1031.93} \AA\ line in our analysis.

We have listed derived equivalent widths and column densities for O~{\small VI} in Table \ref{Tab1}. The 1$\sigma$ error in the equivalent width and column density has been derived using the uncertainty in the {\it FUSE} data and the fitting procedures. Another major uncertainty is due to the overlap of high velocity MW O~{\small VI} with that of the LMC absorption. The separation is distinct for some sightlines viz. Sk-65D21, Sk-67D69, Sk-65D105, HV2543, Sk-66D100, HV5936, Sk-67D211, Sk-69D220, Sk-66D172, Sk-68D137 and D301-1005. For overlapping lines, the lower velocity limit of the LMC O~{\small VI} absorption from \citet{Pathak et al.11} is taken as the upper limit of the MW O~{\small VI} profile. The range of the velocity considered for the integration is presented in Table \ref{Tab1}. The LMC results from \citet{Pathak et al.11} are also presented. The uncertainty in exactly resolving the velocity limits may lead to significant errors in the measurements of high velocity clouds (HVCs).

 For understanding the distribution of O~{\small VI}, we have tried to separate the O~{\small VI} in the MW disk from the IVCs and HVCs. For this the velocity limits are chosen as $|v|\le$ 50 km s$^{-1}$ for the O~{\small VI} in the MW disk, 50 to 100 km s$^{-1}$ for the O~{\small VI} in the IVCs and  $|v|\ge$ 100 km s$^{-1}$ for the O~{\small VI} in the HVCs. The results from these measurements are presented in Table \ref{Tab2}. 
 

\section{Characteristics of O~{\small VI} in the MW}

\subsection{Fe~{\small II} absorption and kinematic properties of O~{\small VI}}

\begin{figure*}
\includegraphics[width=\textwidth]{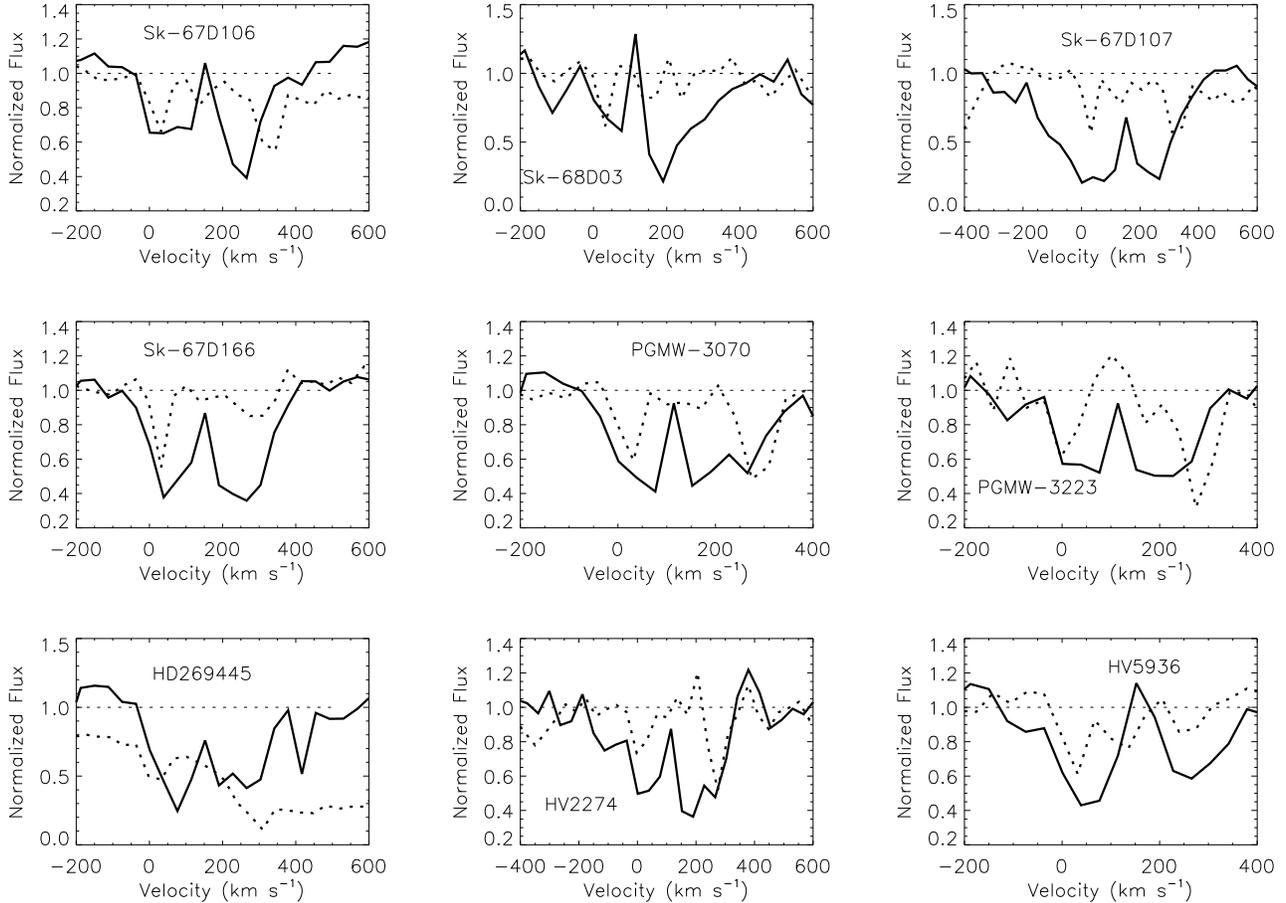}
\caption{O~{\small VI} absorption profiles (black lines) for few lines of sight along with Fe~{\small II} components (dotted lines) in the MW and LMC. Significant absorption (probably stellar) for HD269445 (lower left panel) results in a shift of the continuum of the Fe~{\small II} profile.} 
\label{Fig2}
\end{figure*}

{It is difficult to directly compare the kinematics of O VI in the MW and the LMC based only on the absorption profiles because the O~{\small VI} profiles are broad (Figure \ref{Fig2}) suggesting that they arise in an extended halo of hot gas. A better probe of the kinematics of the gas is the optically thin Fe II absorption line at 1125.448 \citep{Howk02b}. These lines trace the low-ionization gas associated with the relatively high column density warm neutral medium (WNM) and warm ionized medium (WIM) and we observe two components, one at low negative velocities arising in the MW and the other at high positive velocities arising in the LMC (Figure \ref{Fig2};\citep{Howk02b}).}

{The O VI absorption in the MW is observed as a broad line at negative velocities (except for Sk-70D97, BI253 and HV982) arising in both the low velocity components tracked by the Fe II absorption line and in IVC and HVC components where no Fe II is seen.}

Evidence of IVC and/or HVC components in the O~{\small VI} absorption profiles for LMC can also be seen for all sightlines. In comparison to Fe~{\small II}, O~{\small VI} absorption profiles in the LMC extend to the lower velocities. Almost all  Fe~{\small II} profiles are aligned to the high velocity edge of O~{\small VI}. Therefore, Fe~{\small II} present in the LMC is close to the disk and O~{\small VI} is an extended gas.

\subsection{High velocity O~{\small VI}}

Absorption or emission spectra against bright extragalactic sources provide information about the large gaseous structures surrounding the MW. Gas with radial velocities $|V_{LSR}|>100$ km s$^{-1}$ are usually the HVC and IVC is considered to be the gas having velocities  $50<|V_{LSR}|<100$ km s$^{-1}$. HVCs are often described in terms of the H{\small I} 21 cm emission and connect the inner region of a galaxy with the surrounding intergalactic medium. Towards several HVCs, H$\alpha$ emission has been detected \citep{Tufte98,Bland98,Weiner96}. UV spectra of bright extragalactic stars have revealed neutral, weakly, and highly ionized gas at high velocities \citep{Savage81,Welty99,Lehner01,Danforth02,Hoopes,Howk02b,Lehner02}. The HVC has a complex multi-phase structure but the Hubble Space Telescope  and {\it FUSE} results have shown that some HVCs are almost fully ionized \citep{Sembach95,Sembach99,Putman03,Lehner09}. 

Most of the large IVCs and HVCs are located in the inner Galactic
halo at distances d $\leq$ 20 kpc \citep{Wakker07, 2008,Thom06, 2008} except the ``Magellanic Stream'' (MS) which is at $\sim$50 kpc \citep{Gardiner96}. 
The accretion of gas from the two Magellanic Clouds by the MW has been considered as the origin of Magellanic Stream \citep{Wannier72,Fox10,2013,2014,Richter13}.

Other HVCs possibly represent metal-deficient gas that is infalling from the inter-galactic medium. Earlier a Galactic origin of the HVC towards the LMC was proposed \citep{Savage81,deBoer90,Richter99} while recent studies reveal energetic outflow from the LMC as the origin of the HVC complex \citep{Lehner09}.

\subsection{Doppler parameter}
The width of an absorption line is a measure of the total velocity distribution in the gas which, in the case of Galactic lines, may be due to thermal motion, turbulence or multiple components in the line of sight and is characterized by the Doppler parameter ($\it b$) where {$\it b$ = FWHM/2$\sqrt{(LN(2)}$)}.
For high resolution observations such as those with $\it{FUSE}$ where the O~{\small VI} absorption line is fully resolved, the AOD method provides a direct measure of $\it b$. {The b-value has further been corrected by subtracting the second moment for the
smoothed line-spread function from the second moments of the observed profiles.}
\begin{figure*}
\centering
\includegraphics[angle=270, width=12cm]{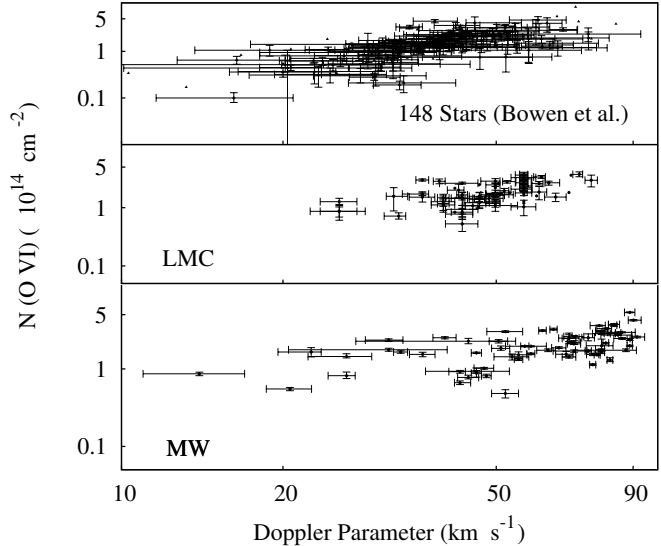}
\caption{O~{\small VI} column density [N(O~{\small VI}] vs Doppler parameter, b. Top panel shows the results obtained by Bowen et al.\citet{2008}. The lower panel represents our results derived for the 69 sightlines and the middle panel represents results for the same 69 sightlines for LMC obtained by Pathak et al. \citet{2011} }
\label{Fig3}
\end{figure*}
The production of O~{\small VI} in collisional ionization equilibrium (CIE) is maximum at 2.8$\times$10$^{5}$ K \citep{Sutherland1993}.
At this temperature the thermal Doppler line width should be 17.1 km s$^{-1}$ corresponding to a FWHM of 28.5 km s$^{-1}$. The observed $\it b$ value for all sightlines of this study ranges from {14.0$^{+1}_{-1}$ km s$^{-1}$ for BREY22 to 91.6$^{+2}_{-2}$ km s$^{-1}$ for HV5936. The corresponding median, average and standard deviation of Doppler parameter are 68.4 km s$^{-1}$, 62.7 km s$^{-1}$ and 20.0 km s$^{-1}$ respectively}.  The average $\it b$-value is larger than expected from thermal broadening in a gas at 2.8$\times$10$^{5}$ K. This may be due to different environments dominated by inflow, outflow and turbulent motions. 

For the MW O~{\small VI} Savage et al. \citet{2003} found $\langle  b \rangle$=61$\pm15$km s$^{-1}$ with a median value of 59 km s$^{-1}$ for 100 extragalactic sightlines. Earlier Jenkins\citet{1978a} found a variation in the value of $\it b$ for the Galactic disk from 10.7 to 56 km s$^{-1}$ with median value of 27 km s$^{-1}$ for O~{\small VI} absorption towards 62 stars observed by the $\it{Copernicus}$ satellite. O~{\small VI} velocity dispersions from 33 to 78 km s$^{-1}$ with an average $\langle{\it b}\rangle =45\pm{11}$ km s$^{-1}$ was reported for 22 halo stars by \citet{Zsargo2003}. Lehner et al. \citet{2011} also derived the O~{\small VI} $\it b$-distribution in the Milky Way disk and found a variation from 20 to 60 km s$^{-1}$.

The distribution of the Doppler parameter and the column density is used to understand the variation in the temperature and the density of the ISM. A linear correlation between N(O~{\small VI}) and $\it b$ was first discussed by Heckman et al.\citet{2002}, who studied the trend using extragalactic sightlines and data from early investigations of O~{\small VI} in the Galaxy.
The discussed correlation between O~{\small VI} column density and Doppler parameter is expected in radiatively cooled or  conductively heated gas. This correlation extends from the local interstellar medium (LISM) to the Galactic disk and the halo and beyond \citep{Heckman et al.2002,Savage03}. Bowen et al.\citet{2008} found a similar relation in the Galactic disk for O~{\small VI}. A good correlation between the Doppler parameter and the O~{\small VI} column density for $\it b$ $>$15 km s$^{-1}$ was reported by Lehner et al.\citet{2011}. In Figure \ref{Fig3} (bottom panel) we plot N(O~{\small VI}) against Doppler parameter $\it b$. We find a correlation as reported in earlier studies. For comparison we also plot the results of Pathak et al.\citet{2011} and Bowen et al.\citet{2008} in the middle and upper panels respectively. O~{\small VI} Doppler line width in the LMC is found to be confined in a narrower range compared to that of the MW. Heckman et al.\citet{2002} suggested that the correlation depends on the characteristic velocity which may be explained by the laws of heating or cooling. The correlation suggests a collision dominated ionisation for O~{\small VI} production \citep{Lehner2011}. In Figure \ref{Fig4}, the histogram of the values of $\it b$ for all the sightlines are shown. In the right panel of Figure \ref{Fig4} we also present the histogram of $\it b$ values as obtained by Pathak et al. \citet{2011} for the LMC. The histogram plots indicate that the $\it b$ values are larger then the value for which maximum O~{\small VI} is expected. Thus, the gas experiences nonthermal motion that dominates the broadening. There may also be undefined components contributing to the profile width \citep{Lehner2011}.

\subsection{Column density}

 {We find a variation of about an order of magnitude in the O~{\small VI} column densities in the MW lying between 13.68 to 14.73 (in units of atoms cm$^{-2}$) for log N(O~{\small VI}). The mean MW O~{\small VI} column density is found to be {14.26} atoms cm$^{-2}$ and the median value for our sample is {14.27} atoms cm$^{-2}$. The projections of the total column densities perpendicular to the Galactic plane, log (N sin $|b|$) are in approximately the same range: from 13.42 to 14.50 atoms cm$^{-2}$ with average and median values of {14.00 and 14.02} atoms cm$^{-2}$ respectively. Similar values of O~{\small VI} column densities are reported in different literature as given in Table \ref{Tab4}.
\begin{figure*}
\centering
\includegraphics[angle=270, width=8.4cm]{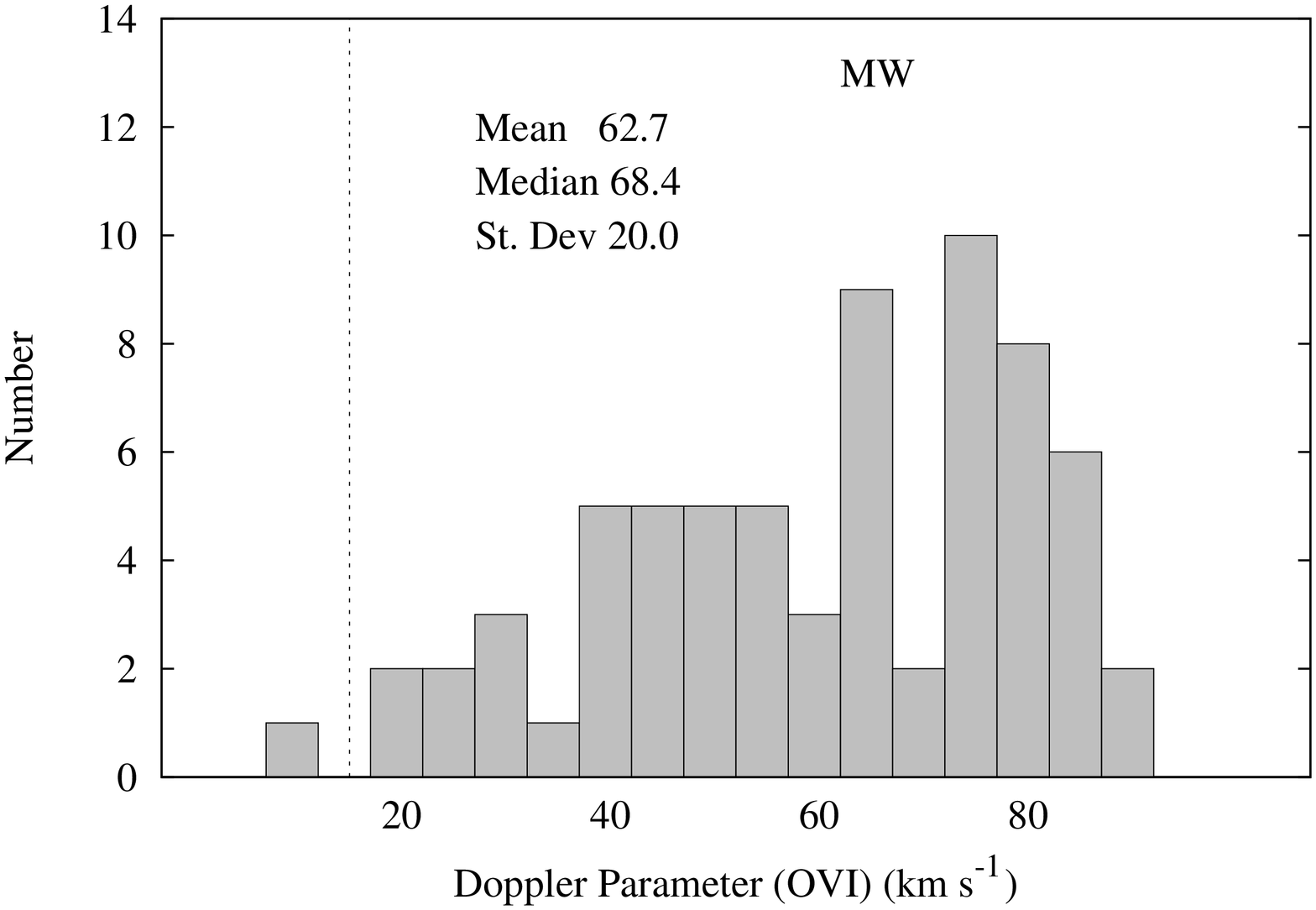}
\includegraphics[angle=270, width=8.4cm]{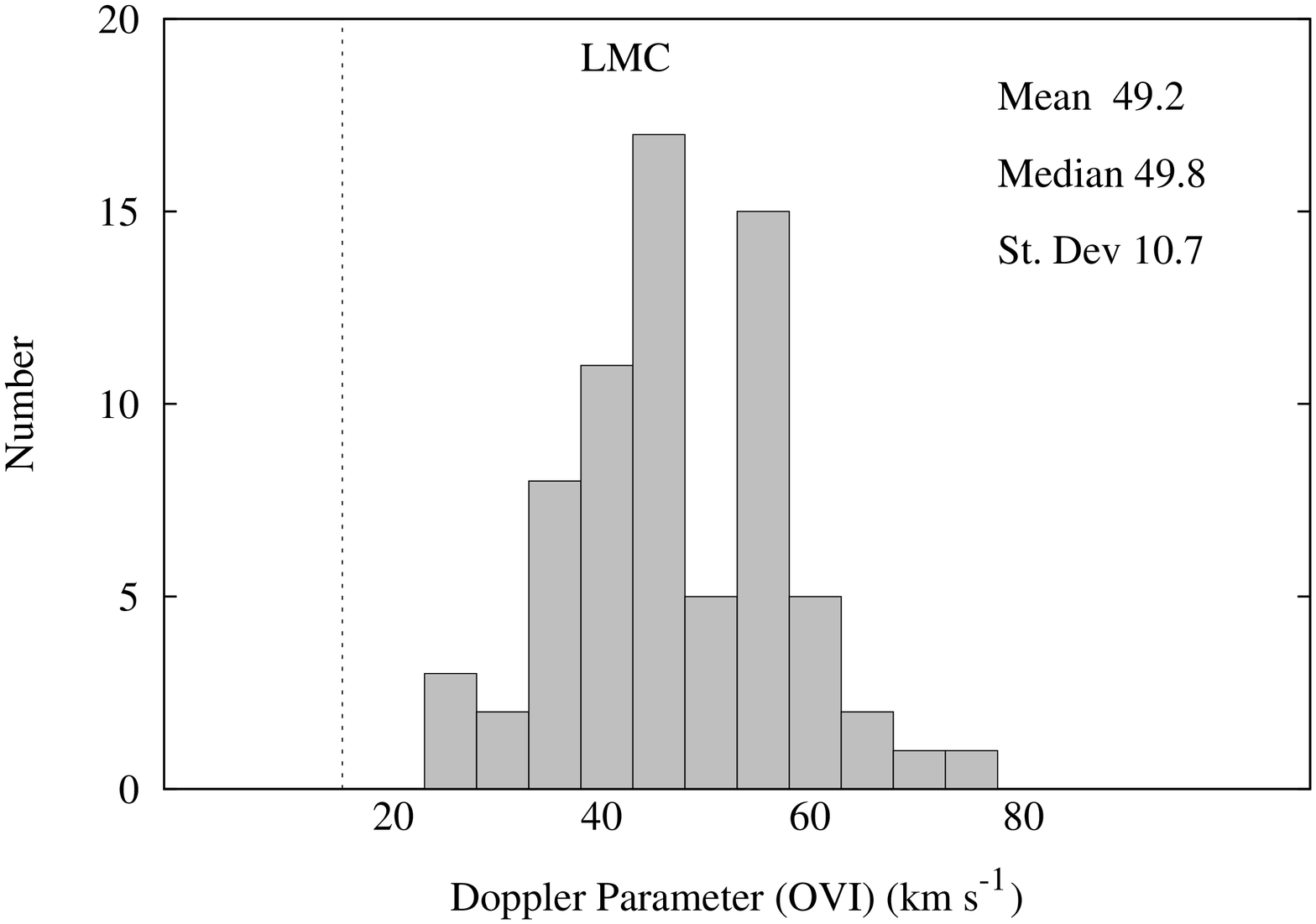}
\caption{Doppler parameter distribution of O~{\small VI} considering a bin size of 5 km s$^{-1}$ for the MW (left) and the LMC (right). The vertical dotted line represents the expected b-value at maximum O~{\small VI} abundance temperature in CIE. }
\label{Fig4}
\end{figure*}

{As discussed in section 4.1, the O~{\small VI} profiles for all sightlines in our sample have HVC and/or IVC components. For the low velocity component, we have derived the MW O~{\small VI} column densities by integrating over $|v|\le$50 km s$^{-1}$ (Table \ref{Tab2}). This velocity range is nearly free from the IVC. The log N(O~{\small VI}) for this velocity range varies from 12.55 to 14.52 atoms cm$^{-2}$. The mean column density log N(O~{\small VI}) for this range is 13.94 atoms cm$^{-2}$.

 The log N(O~{\small VI}) for the velocity range $|v|\le$ 100 km s$^{-1}$ representing the disk with the IVC component, varies from  14.63 to {13.61} atoms cm$^{-2}$. The mean column density log N(O~{\small VI}) for this range is {14.24} atoms cm$^{-2}$. There are {5} sightlines where the IVC contribution to the total O~{\small VI} column density is greater than  $70 \%$ and 14 sightlines where the contribution is $> 50 \%$. Howk et al.\citet{2002b} report O~{\small VI} column density values in the Galactic halo from 13.61 to 14.23 atoms cm$^{-2}$ in the velocity range -50 to +50 km s$^{-1}$ and 14.22 to 14.67 atoms cm$^{-2}$ when integrated over total velocity range for 12 stars in the LMC. The individual sightline of the 12 stars have varying column density values for the two corresponding velocity limits.}

Out of the 69 sightlines, we have not found HVC component in case of BREY22 and HDE269599 (Figure \ref{Fig1}). Except these two, it is observed that the HVC contribution is significant for all other sightlines. For $|v|> $100~ km~ s$^{-1}$ region, log N(O~{\small VI}) ranges from 9.78 to 14.03 atoms cm$^{-2}$ with a mean of {12.87} atoms cm$^{-2}$. There are {9} cases where the HVC contribution to the total column density is $\ge 20 \%$. The highest contribution from this velocity range is found in case of Sk-71D45.}

\subsection{Comparison with the LMC and the SMC}
{The metallicity of the MW is higher than that of the LMC and the SMC with SMC being the lowest. This implies that the O~{\small VI} abundance in the MW would be higher compared to that of the LMC and the SMC. O~{\small VI} column density for all the 69 lines in the LMC have already been reported by Pathak et al.\citet{2011}. They found high abundance of O~{\small VI} with log N(O~{\small VI}) in the range of 13.72 to 14.57 atoms cm$^{-2}$ and a mean of 14.23 atoms cm$^{-2}$. The variation of column density in the LMC is lower than the MW. The mean O~{\small VI} column density for the MW is higher then the LMC. Considering the inclination angle of the LMC to be 33$^\circ$, Pathak et al. \citet{2011} calculated the projected O~{\small VI} column density on the plane as 14.16. Table \ref{Tab1} presents a comparison of the equivalent widths and the column densities in the two galaxies for all sightlines.  We find that the O~{\small VI} column densities in the MW are higher (or comparable) to the LMC values. The O~{\small VI} absorption profiles are similar to that of the LMC which is in agreement with Pathak et al.\citet{2011}.}

Howk et al. \citet{2002a} analysed the distribution and kinematics of O~{\small VI} absorption towards 12 early-type stars in the LMC. They observed O~{\small VI} absorption for all 12 stars and derived column densities in the range of 13.9--14.6 atoms cm$^{-2}$ with a mean of 14.37 atoms cm$^{-2}$. They report that the average column density of O~{\small VI} and the dispersion of the individual measurements about the mean are identical to those measured for the halo of the MW.  

Since, the O~{\small VI} absorption is very patchy in nature and the O~{\small VI} abundance depends on the local ISM conditions, it is extremely difficult to compare the O~{\small VI} column densities of the MW and the LMC. 

O~{\small VI} absorption in the SMC was surveyed by \citet{Hoopes} for 18 early-type stars. They report a widespread presence of O~{\small VI} with a mean value of N(O~{\small VI}) as 14.53 atoms cm$^{-2}$. While the metallicity of the SMC is lower, the mean column density is higher than the MW and the LMC. Highest abundance of O~{\small VI} has been found for NGC 346 which is a star forming region of the SMC. In the SMC the O~{\small VI} profile is shifted to higher velocities compared to the lower ionization gas traced by Fe~{\small II} absorption \citep{Hoopes}.

\subsection{O~\small{VI} column density and angular variation}
{The O~{\small VI} column densities have been measured using AOD method for 69 stars in the LMC [(l,b)=280$\degr$.5, - 32$\degr$.9 and d= 50 kpc]. As evident from Table \ref{Tab1}, there exists significant variation in the N(O~{\small VI}) values. The lowest value of column density has been found to be 4.82$\times 10^{13}$ atoms cm$^{-2}$ for SK-65D63. For LH91486 the value is highest which is 5.36$\times 10^{14}$ atoms cm$^{-2}$. This gives a variation of about one order of magnitude.}

{The column density variation with respect to the angular scale may provide us information on physical properties of the regions in which O VI is produced. An idea of the physical dimension and shape of the O~{\small VI} bearing clouds may be obtained from the amount of change in N(O~{\small VI}) and angular scale variations. Though uniform distribution of hot gas in the galactic halo was considered \citep{Spitzer1956}, soon  it became clear that the O~{\small VI} distribution is not smooth in the galaxy \citep{Savage03}. Howk et al. \citet{2002b} found large variations in the MW O~{\small VI} column densities over small angular scales towards 12 LMC and 11 SMC stars. They found column density variations for the smallest scales with a value of $\bigtriangleup \theta \sim$ 1$\arcmin$.8. No significant N(O~{\small VI}) variation was found towards 4 stars in NGC 6752 that are separated by 2$\arcmin$.2 - 8$\arcmin$.9 \citep{Lehner04}.}

\begin{figure*}
\centering
\includegraphics[angle=270,width=10cm]{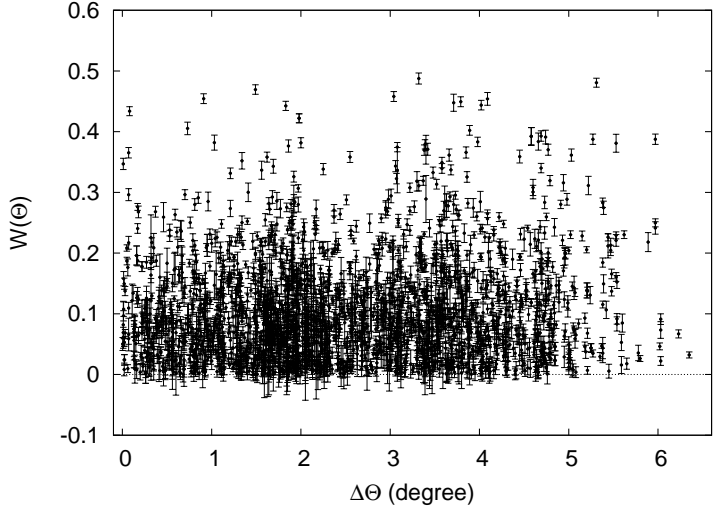}
\caption {Absolute logarithmic column density difference, $\Delta$N(O~{\small VI}) = N(O~{\small VI})$_{i}$--N(O~{\small VI})$_{j}$ for each (ij) pair of sightlines as a function of angular separation between the pairs. }

\label{Fig6}
\end{figure*}
{The sightlines in our study are separated by angular distances of 0.0025$\degr \le \bigtriangleup \theta \le$ 6.35$ \degr$. The smallest scale for which O~{\small VI} column density variation has been measured is $\bigtriangleup \theta$ $\sim$ 0.0025$\degr$ for the closest pairing stars Sk-69D243 and MK42. On the other hand Sk-65D21 and Sk-68D03 are found to be the most distant lines of sight ($\bigtriangleup \theta$ $\sim$ 6.35$\degr$). Assuming O~{\small VI} absorption by HVCs located in the halo of the Galaxy towards the LMC at a distance of 40 kpc from the Sun (\citet{Lehner09, Sarma et al.2014}), we find the distance between the closest and most distant stars to be 1.75 pc and 4.43 kpc respectively.}
\begin{figure*}
\centering
\includegraphics[angle=270,width=10cm]{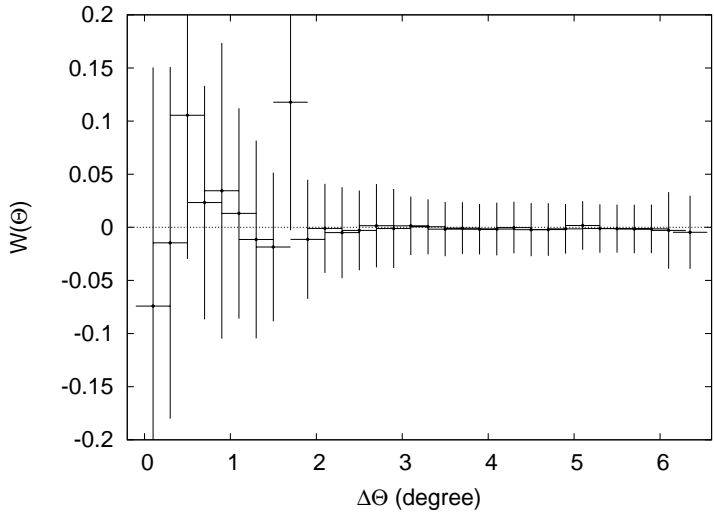}
\caption {Autocorrelation function, w$(\theta)$ vs Angular separation for the 69 sightlines considering 0.2$\degr$ bin spacing.{We consider the binsize and standard deviation as the error for $\Delta(\theta)$ and w$(\theta)$ respectively. }
}
\label{Fig5}
\end{figure*}
{Figure \ref{Fig6}~ shows the difference in column densities for each pair of sightlines versus the angular separation of the pair. We measure absolute column density difference i.e. $\Delta$N(O~{\small VI}) = N(O~{\small VI})$_{i}$--N(O~{\small VI})$_{j}$ for each pair of sightlines i and j. In Figure \ref{Fig6} we show relative variation in $\Delta$N(O~{\small VI}) for object pairs plotted against the angular separation of the pair. 
There exists clear variation in the amount of N(O~{\small VI}) over all angular scales. The average and standard deviation of logarithmic values of column density difference of each sightline is found to be 13.83 and {0.49} atoms cm$^{-2}$. This result confirms the patchiness of O~{\small VI} in the Galaxy.}

{Following Howk et. al \citet{2002b}, we also derived the angular autocorrelation function (ACF) w($\theta$) for O~{\small VI} column densities towards all the 69 sightlines. The ACF is calculated using the expression 
 \begin{equation}
 { {w}(\theta)= \frac{\Sigma_i\Sigma_{j{\neq}i}(N_i - \langle N \rangle)(N_j - \langle N \rangle)(\sigma_i \sigma_j)^{-1}}{\langle N \rangle^2\Sigma_i \Sigma_{j{\neq}i}(\sigma_i \sigma_j)^{-1}}},
\end{equation}
{ where N$_i$ and N$_j$ are the O~{\small VI} column densities with uncertainties $\sigma_i$ and  $\sigma_j$. {This is the same expression as Howk et. al \citet{2002b} used except that the weight factors have been raised to (-1).}

The average column density of 69 sightlines is represented by $\langle N \rangle$. We have considered the bin spacing as  0.2$\degr$ for the whole data set. The derived values of w($\theta$) are shown in Figure \ref{Fig5}. }

{We find significant variation of w($\theta$) on small scales. Below 2$\degr$, the maximum factor by which the ACF varies is $\sim$ {1.6}. This indicates that the cloud structure may be much smaller. At large scales, the variation of w($\theta$) is not significant ($\sim$ 0) which is similar to the earlier result \citep{Howk02b}.}

{Howk et al.\citet{2002b} discuss the angular correlation function of the O~{\small VI} measurements, finding no preferred angular scale for variations. Savage et al.\citet{2003} showed that the variation seen towards the LMC and the SMC extend to larger angular scales. They favoured models in which the O~{\small VI}-bearing medium is composed of small complex cloud-like or sheet-like distribution of material \citep{Howk02b, Savage03}.}

\subsection{Scale Height}
Another aim of this work is to study the distribution of O~{\small VI} in the Galactic plane with a large sample size compared to earlier studies (\citet{Jenkins 1978b, Savage03, Bowen et al., Savage2009}). We assume an exponential gas distribution with O~{\small VI} volume density as a function of height (z),  [n($|z|$)=n$_{0}$ exp (-$|z|$/h) where h is scale height and n$_{0}$ is midplane density]. The column density N(x) perpendicular to the plane for an object with latitude b is given by N(x) sin $|b|$=n$_{0}$h[1-exp (-$|z|$/h)], where N(x) sin $|b|$ is the line-of-sight column density of an object. For extra-galactic objects ($|z| \gg$ h), N(x) sin $|b|$=n$_{0}$h is used.

Using {\it Copernicus} data the midplane density n$_{0}$ and corresponding O~{\small VI} scale height have been estimated to be 2.8$\times$10$^{-8}$ cm$^{-3}$ and 0.3 kpc respectively by \citet{Jenkins 1978b}. The {\it Copernicus} O~{\small VI} observations have been reanalyzed by Shelton \& Cox \citet{1994} and they found that the midplane O~{\small VI} density beyond the Local Bubble (LB) is (1.3 - 1.5)$\times$10$^{-8}$ cm$^{-3}$ for an O~{\small VI} absorbing layer with h$\sim$ 3 kpc. Savage et al.\citet{2000} estimate n$_{0}$ = 2.0$\times$10$^{-8}$ cm$^{-3}$ from {\it Copernicus} O~{\small VI} survey and derived scale height h$>$2 kpc. Using $\it{FUSE}$ data of 100 extragalactic objects Savage et al.\citet{2003} derived a scale height of h$\sim$2.5 kpc. \citet{Bowen et al.} used $\it{FUSE}$ data of 148 early type stars and estimated midplane density n$_{0}$ = 1.33$\times$10$^{-8}$ cm $^{-3}$ which decreases away from the plane of the Galaxy.  \citet{Bowen et al.} considered a correction for O~{\small VI} absorption in the LB and measured O~{\small VI} scale height as 3.2 kpc at negative
latitudes or 4.6 kpc at positive latitudes.

We measure the scale height for all the sightlines of our study. For a midplane density n$_{0}$=1.64$\times$10$^{-8}$ cm$^{-3}$ \citep{Savage2009}, we obtain the scale height using the simple relation h=N(x) sin $|b|$/n$_{0}$. The value ranges from { 0.53 to 6.4 kpc. The average values of the O~{\small VI} scale height is 2.28$\pm$1.06 kpc}. Savage \& Wakker \citet{2009} using a large object sample (109 stars and 30 extragalactic objects) derived the O~{\small VI} scale height to be 2.6$\pm$0.5 kpc from the Galactic plane. We have also calculated the scale height for all the sightlines for the column density derived in the range from minimum to 100 kms$^{-1}$. The average scale height in this case has been found to be {2.05} kpc.

From the O~{\small VI} scale height value, the temperature of the ionized plasma can be found as estimated by Savage \& Wakker \citet{2009}. Considering hydrostatic equilibrium at temperature T for an isothermal gas, the scale height is given by h=kT/$\langle{m}\rangle g(\vert{z}\vert)$. The value of gravitational acceleration towards the disk g($\vert{z}\vert)$, in the solar neighbourhood for a distance 1-10 kpc is $\sim 10^{-8}$cm s$^{-2}$ \citep{Kalberla 2008}. Considering average mass per particle $\langle{m}\rangle$=0.73m$_{H}$, for a scale height {h=2.28} kpc the temperature of the gas is found to be T$\sim$ {0.62}$\times$ 10$^{6}$ K. This temperature is $\sim$ {2.21} times larger than the transition temperature of O~{\small VI}. Savage \& Wakker \citet{2009} derived this temperature as $\sim$ 0.8$\times$ 10$^{6}$ K for a scale height value h=3 kpc. 

\section{Summary and conclusion}

We have studied the properties of O~{\small VI} in the MW along the lines of sight towards 69 stars in the LMC using {\it FUSE} spectra. The observed absorption lines are studied using AOD method which reveal significant variation in O~{\small VI} column densities over small angular scale. The important results of these study may be summarized as follows:

1. We find O~{\small VI} absorption in the MW with IVC and HVC components. The highest column density measured for the MW is log N(O{\small VI}) = 14.73~atoms~cm$^{-2}$ and the minimum value is log N(O{\small VI}) = 13.68~atoms~cm$^{-2}$. The mean MW O~{\small VI} column density is found to be {14.26} atoms cm$^{-2}$. The median value of our sample is {14.27} atoms cm$^{-2}$. The logarithm of the column densities perpendicular to the Galactic plane varies between 13.42 and 14.50 atoms cm$^{-2}$ with an average value of {14.00} atoms cm$^{-2}$. We observed the O~{\small VI} absorption line at 1036 \AA\ but did not use it as it overlaps with the CII* absorption line.

2. There is a significant variation in the O~{\small VI} column densities on all scales (0.0025${\degr}$ - 6.35${\degr}$). Higher variation towards smaller angular scale indicates that the O~{\small VI} bearing clouds are small in size.

3. The measured O~{\small VI} column densities can be described by a patchy exponential distribution in the MW which is in accordance with earlier measurements of O~{\small VI} absorption for the Galaxy. 

4.  The velocity dispersion ($\it b$-values) of the O~{\small VI} absorption profiles range from {14.0 to 91.6 km s$^{-1}$. The median, average and standard deviation of equivalent width are 68.4, 62.7 and 20.0 km s$^{-1}$ respectively}. The average b-value is larger than expected from thermal broadening in gas at 2.8$\times$10$^{5}$ K. This may be due to different environments dominated by inflow, outflow and turbulent motions. 

5. The broad ($>$25.24 km s$^{-1}$) O~{\small VI} absorption profiles suggest collisional ionization at the interface of warm-hot ISM to be the mechanism that produces this ion.

6. The O~{\small VI} column density and the Doppler parameter, $\it b$, are found to be correlated for our sample of sightlines. This  confirms earlier results.

7. A kinematical comparison of Fe~{\small II} and O~{\small VI} line profiles reveal the presence of HVC and/or IVC components in the O~{\small VI} absorption along all sightlines. In the Fe~{\small II} profiles we have not observed either of these components.The distribution of O~{\small VI} is significantly different than that of Fe~{\small II}. The broad absorption profiles of O~{\small VI}  trace extended layers compared to that of the Fe~{\small II}-bearing layer.

8.  For a midplane density n$_{0}$=1.64$\times$10$^{-8}$ cm$^{-3}$, we measured the O~{\small VI} scale height as {2.28$\pm$1.06} kpc. This is in accordance with earlier results.

A thorough mapping of O~{\small VI} in the MW and in nearby galaxies will be helpful in understanding the distribution and kinematics of the ionized gas. Future scientific mission which may cover O~{\small VI} wavelengths will definitely explore much more about the hot ionized phase of the ISM. 
\clearpage
\onecolumn 
{
\scriptsize
\begin{longtable}{lcccccc}
\caption{O~{\tiny VI} column densities, equivalent widths and the corresponding velocity limits in the Milky Way and the LMC.}
\label{Tab1}\\
\hline \hline\\[-2ex]

Target Name &   \multicolumn{3}{c}{Milky Way}&  \multicolumn{3}{c}{LMC} \\

& Limit&EW&log N(O~{\scriptsize VI})& Limit &EW&log N(O{\scriptsize VI}) \\

&(km~s$^{-1}$)&(m\AA)&(dex)&(km~s$^{-1}$)&(m\AA)&(dex) \\
\endfirsthead

\multicolumn{6}{c}{{\tablename} \thetable{} -- Continued} \\[0.5ex]
\hline \hline\\[-2ex]

Target Name &   \multicolumn{3}{c}{Milky Way}&  \multicolumn{3}{c}{LMC} \\

& Limit&EW&log N(O~{\scriptsize VI}) &Limit &EW &log N(O{\scriptsize VI}) \\

&(km~s$^{-1}$)&(m\AA)&(dex)&(km~s$^{-1}$)&(m\AA)&(dex) \\
\hline
\endhead
\endfoot
\endlastfoot
\hline
SK-67D05	&-30,175	& 259.9$\pm$15		&14.26$^{+0.03}_{-0.03}$	&175, 330	&73$\pm$6	&13.72$^{+0.10}_{-0.13}$         \\
SK-68D03	&-20,120	& 122$\pm$18		&14.06$^{+0.04}_{-0.04}$	&180,330	&112$\pm$8	&14.02$^{+0.11}_{-0.15}$         \\
BI13		&-35,120	& 190.0$\pm$27	        &14.25$^{+0.06}_{-0.08}$	&175, 315	&88$\pm$4	&13.94$^{+0.08}_{-0.09}$         \\
SK-67D18	&-30,135	& 130.3$\pm$17	        &14.12$^{+0.13}_{-0.16}$	&165, 330	&132$\pm$7	&14.16$^{+0.09}_{-0.13}$         \\
Sk-67D20	&-30,175	& 200.2$\pm$3	        &14.19$^{+0.06}_{-0.07}$	&175, 335	&147$\pm$8	&14.08$^{+0.11}_{-0.16}$         \\
PGMW-3070	&-50,120	& 139.2$\pm$15	        &14.23$^{+0.12}_{-0.16}$	&180, 345	&132$\pm$23     &14.10$^{+0.13}_{-0.18}$         \\
LH103102	&-20,150	& 287.0$\pm$23	        &14.38$^{+0.02}_{-0.03}$	&180, 330	&132$\pm$23	&14.16$^{+0.03}_{-0.04}$         \\
LH91486		&-30,175        & 459.9$\pm$22		&14.73$^{+0.05}_{-0.06}$	&175, 385	&266$\pm$29	&14.47$^{+0.09}_{-0.11}$         \\
PGMW-3223	&-50,120	& 138.9$\pm$17		&14.01$^{+0.06}_{-0.07}$	&175, 315	&129$\pm$15	&14.10$^{+0.06}_{-0.07}$         \\
HV2241		&-35,165        & 330.0$\pm$66		&14.56$^{+0.05}_{-0.06}$	&165, 365	&271$\pm$18	&14.50$^{+0.04}_{-0.03}$         \\
HD32402		&-40,140        & 253.5$\pm$22		&14.40$^{+0.08}_{-0.09}$	&160, 320	&271$\pm$8	&14.48$^{+0.02}_{-0.02}$         \\
SK-67D32	&-40,165	&{192.0$\pm$28}		&{13.82$^{+0.09}_{-0.11}$}	&165, 360	&129$\pm$7	&14.11$^{+0.03}_{-0.04}$        \\
SK-65D21	&-20,165	& 111.0$\pm$29		&13.91$^{+0.08}_{-0.09}$	&225, 340	&94$\pm$23	&13.94$^{+0.11}_{-0.16}$         \\
HV2274		&-20,125	& 219.0$\pm$20	        &14.19$^{+0.09}_{-0.11}$	&165, 345	&160$\pm$11	&14.14$^{+0.10}_{-0.15}$         \\
Sk-66D51	&-30,170	& 391.8$\pm$21		&14.47$^{+0.03}_{-0.02}$	&170, 370	&185$\pm$10	&14.26$^{+0.01}_{-0.02}$         \\
NGC1818-D1	&-50,150	& 320$\pm$20		&14.46$^{+0.03}_{-0.04}$	&150, 340	&280$\pm$13	&14.52$^{+0.05}_{-0.05}$        \\
SK-70D69	&-50,150	&{115.0$\pm$56}	        &{13.96$^{+0.13}_{-0.19}$}	&150, 295	&182$\pm$12	&14.18$^{+0.07}_{-0.08}$         \\
Sk-67D69	&-20,170	&{138.0$\pm$18}		&{13.97$^{+0.07}_{-0.10}$}	&170, 340	&232$\pm$28	&14.40$^{+0.05}_{-0.06}$         \\
MACHO78-6097	&-30,165	&236.9$\pm$50		&14.32$^{+0.05}_{-0.05}$	&165, 320 	&151$\pm$25	&14.15$^{+0.05}_{-0.05}$         \\
BI130		&-30,165	&{339.9$\pm$6}		&{14.39$^{+0.11}_{-0.12}$}	&165, 320	&94$\pm$7	&13.89$^{+0.07}_{-0.09}$         \\
SK-69D94	&-20,160	& 149.8$\pm$8		&14.22$^{+0.06}_{-0.06}$	&160, 340	&150$\pm$3	&14.22$^{+0.03}_{-0.03}$         \\
SNR0519-697	&2,160		& 339.0$\pm$16		&14.44$^{+0.05}_{-0.06}$	&160, 300	&97$\pm$4	&13.97$^{+0.07}_{-0.08}$         \\
BREY22		&10,160         & 105.6$\pm$20		&13.94$^{+0.08}_{-0.10}$	&160, 330	&126$\pm$24	&14.07$^{+0.06}_{-0.07}$        \\
HD269445	&-30,175	& 301.2$\pm$31		&14.49$^{+0.02}_{-0.02}$	&175, 365	&200$\pm$10	&14.29$^{+0.04}_{-0.05}$         \\
SK-69D124	&-35,190	& 136.4$\pm$26		&13.95$^{+0.10}_{-0.10}$	&190, 430	&331$\pm$8	&14.57$^{+0.04}_{-0.05}$         \\
SK-67D105	&-40,180	& 311.3$\pm$40		&14.42$^{+0.05}_{-0.05}$	&180, 320	&110$\pm$8	&13.92$^{+0.14}_{-0.23}$         \\
SK-67D106	&-20,150	& 113.0$\pm$22		&13.89$^{+0.08}_{-0.09}$	&175, 345	&180$\pm$7	&14.30$^{+0.10}_{-0.13}$         \\
SK-67D107	&-40,120	& 427.6$\pm$2		&14.63$^{+0.03}_{-0.03}$	&160, 360	&254$\pm$10	&14.45$^{+0.04}_{-0.05}$        \\
HD36521		&-30,175	& 190.0$\pm$3		&14.24$^{+0.13}_{-0.19}$	&175, 340	&126$\pm$9	&14.07$^{+0.09}_{-0.12}$         \\
SK-68D82	&-20,165	& 226.6$\pm$12		&14.30$^{+0.08}_{-0.09}$	&165, 320	&141$\pm$10	&14.17$^{+0.04}_{-0.04}$        \\
BI170		&-35,165        &{190.0$\pm$59}		&{14.23$^{+0.03}_{-0.03}$}	&165, 365	&235$\pm$20	&14.43$^{+0.02}_{-0.03}$         \\
SK-67D111	&-30,175	& 260.0$\pm$10		&14.46$^{+0.05}_{-0.05}$	&175, 365	&214$\pm$19	&14.34$^{+0.05}_{-0.05}$         \\
HV2543		&-35,160        & 255.6$\pm$25		&14.30$^{+0.10}_{-0.09}$	&160, 365	&156$\pm$42	&14.23$^{+0.04}_{-0.03}$         \\
SK-70D91	&-45,160	& 243.0$\pm$20		&14.39$^{+0.04}_{-0.04}$	&160, 365	&256$\pm$7	&14.43$^{+0.04}_{-0.03}$         \\
Sk-66D100	&-35,160	&{222.0$\pm$64}		&{14.21$^{+0.04}_{-0.05}$}	&160, 340	&214$\pm$21	&14.34$^{+0.03}_{-0.03}$         \\
HDE269599	&-40,100	& 169.6$\pm$21		&14.18$^{+0.04}_{-0.05}$	&165, 320	&123$\pm$6	&14.05$^{+0.03}_{-0.03}$         \\
SK-65D63	&-40,150	& 54.0$\pm$19		&13.68$^{+0.16}_{-0.26}$	&180, 375	&184$\pm$8	&14.21$^{+0.04}_{-0.04}$         \\
HV982		&40,175         &117.5$\pm$55		&13.91$^{+0.09}_{-0.09}$	&175, 360	&208$\pm$58	&14.33$^{+0.12}_{-0.17}$         \\
SK-70D97	&20,175		& 74.9$\pm$12		&13.74$^{+0.10}_{-0.12}$	&175, 375	&226$\pm$4	&14.39$^{+0.03}_{-0.04}$         \\
Sk-67D144	&-30,165	& 185.2$\pm$76		&14.19$^{+0.06}_{-0.07}$	&165, 335	&194$\pm$10	&14.25$^{+0.06}_{-0.07}$         \\
BI184		&-30,165        & 295.4$\pm$28		&14.51$^{+0.04}_{-0.04}$	&165, 330	&118$\pm$10	&14.04$^{+0.08}_{-0.10}$         \\
NGC2004-B15	&-30,165	& 200.5$\pm$24		&14.22$^{+0.09}_{-0.12}$	&175, 330	&92$\pm$7	&13.86$^{+0.05}_{-0.07}$         \\
SK-71D38	&-50,165	& 350.1$\pm$21	        &14.52$^{+0.04}_{-0.04}$	&165, 315	&94$\pm$7	&13.92$^{+0.07}_{-0.08}$         \\
Sk-71D45	&-30,160	& {203.0$\pm$12}	&{14.19$^{+0.05}_{-0.06}$}	&160, 345	&194$\pm$9	&14.26$^{+0.06}_{-0.06}$         \\
SK-67D166	&-20,165	& 260.0$\pm$21		&14.29$^{+0.04}_{-0.04}$	&165, 390	&206$\pm$9	&14.32$^{+0.05}_{-0.05}$         \\
SK-67D168	&-30,165	& 149.2$\pm$40		&14.27$^{+0.03}_{-0.02}$	&165, 375	&186$\pm$12	&14.26$^{+0.05}_{-0.05}$         \\
SNR0532-675	&-70,165	&{292.0$\pm$80}		&{14.50$^{+0.02}_{-0.02}$}	&165, 345	&147$\pm$9	&14.16$^{+0.07}_{-0.08}$         \\
SK-67D191	&-40,165	& 462.7$\pm$27		&14.64$^{+0.01}_{-0.01}$	&165, 340	&229$\pm$16	&14.42$^{+0.02}_{-0.02}$         \\
HV5936		&-30,175        & 286.0$\pm$25		&14.41$^{+0.04}_{-0.05}$	&175, 375	&174$\pm$8	&14.18$^{+0.07}_{-0.08}$         \\
Sk-69D191	&-30,165	& 219.0$\pm$41		&14.22$^{+0.04}_{-0.04}$	&165, 340	&185$\pm$25	&14.22$^{+0.06}_{-0.07}$         \\
J053441-693139	&-35,165	&189.7$\pm$63		&14.18$^{+0.16}_{-0.24}$	&165, 330	&182$\pm$33	&14.22$^{+0.05}_{-0.05}$         \\
Sk-67D211	&-40,160	&{222.0$\pm$55}		&{14.13$^{+0.05}_{-0.07}$}	&160, 350	&144$\pm$8	&14.11$^{+0.10}_{-0.13}$         \\
BREY64		&10,180         & 239.0$\pm$43		&14.25$^{+0.16}_{-0.25}$	&180, 330	&139$\pm$8	&14.20$^{+0.16}_{-0.23}$         \\
SNR0536-692	&38, 165	&139.7$\pm$39.7		&14.16$_{-0.24}^{+0.56}$	&165, 320	& 116$\pm$7	&14.03$_{+0.05}^{-0.07}$         \\
SK-69D220	&-40,155	& 222.0$\pm$23		&14.33$^{+0.02}_{-0.03}$	&160, 315	&128$\pm$9	&14.09$^{+0.08}_{-0.11}$         \\
Sk-66D172	&-30,175	& 316.9$\pm$38		&14.47$^{+0.03}_{-0.03}$	&175, 360	&173$\pm$9	&14.20$^{+0.05}_{-0.05}$         \\
BI253		&-20,160	& 260.0$\pm$46		&14.36$^{+0.04}_{-0.05}$	&160, 300	&259$\pm$46	&14.46$^{+0.04}_{-0.05}$         \\
SK-68D137	&-20,165	& 307.0$\pm$51		&14.37$^{+0.05}_{-0.06}$	&165, 330	&234$\pm$7	&14.45$^{+0.03}_{-0.02}$         \\
MK42		&-20,160        & 367.4$\pm$12		&14.48$^{+0.09}_{-0.11}$	&160, 330	&228$\pm$24	&14.41$^{+0.07}_{-0.08}$         \\
SK-69D243	&10,150		& 293.2$\pm$15		&14.40$^{+0.08}_{-0.09}$	&150, 345	&307$\pm$15	&14.56$^{+0.05}_{-0.06}$         \\
30DOR-S-R136	&-30,165	& 212.8$\pm$45		&14.16$^{+0.12}_{-0.13}$	&165, 320	&185$\pm$25	&14.25$^{+0.05}_{-0.06}$         \\
SK-69D246	&-40,145	& 358.9$\pm$2		&14.56$^{+0.03}_{-0.03}$	&155, 325	&211$\pm$7	&14.37$^{+0.03}_{-0.03}$         \\
HDE269927	&-40,160	& 228.6$\pm$21		&14.34$^{+0.06}_{-0.07}$	&160, 320	&245$\pm$8	&14.42$^{+0.03}_{-0.04}$        \\
SK-69D257	&-40,160	& 245.8$\pm$98		&14.49$^{+0.05}_{-0.06}$	&160, 315	&171$\pm$8	&14.20$^{+0.07}_{-0.08}$         \\
SNR0543-689	&0, 160		& 238.8$\pm$8.47	&14.24$^{+0.07}_{-0.08}$	&160, 360	&186$\pm$34	&14.27$^{+0.10}_{-0.12}$         \\
D301-1005	&-30,165	& 279.8$\pm$72		&14.47$^{+0.04}_{-0.05}$	&165, 385	&284$\pm$57	&14.53$^{+0.02}_{-0.03}$         \\
SK-67D250	&-40,165	& 212.8$\pm$58		&14.29$^{+0.04}_{-0.05}$	&165, 375	&316$\pm$33	&14.57$^{+0.03}_{-0.03}$         \\
D301-NW8	&-30, 175       &{204.2$\pm$35.5}	&{14.15$^{+0.02}_{-0.02}$}	&175, 365	&228$\pm$30	&14.42$^{-0.02}_{+0.03}$         \\
SK-70D115	&-40,165	& 170.8$\pm$1		&14.36$^{+0.15}_{-0.23}$	&165, 330	&186$\pm$11	&14.23$^{+0.03}_{-0.04}$         \\
\hline
\end{longtable}
\medskip
\noindent
Notes. LMC values are taken from Pathak et al.\citet{2011}
}

{
\scriptsize
\begin{longtable}{lcccccc}
\caption{O~{\tiny VI} column densities with the corresponding velocity limits in the Milky Way.}
\label{Tab2}\\
\hline \hline\\[-2ex]

Target name&limit&log N(O~{\scriptsize VI}) & limit&log N(O~{\scriptsize VI})&log N(O~{\scriptsize VI})$^{\dag}$  \\

&(km~s$^{-1}$)&(dex)&(km~s$^{-1}$)&(dex)& (dex) \\
\endfirsthead

\multicolumn{7}{c}{{\tablename} \thetable{} -- Continued} \\[0.5ex]
\hline \hline\\[-2ex]

Target name &limit&log N(O~{\scriptsize VI})&  limit&log N(O~{\scriptsize VI})&log N(O~{\scriptsize VI})$^{\dag}$  \\

&(km~s$^{-1}$)&(dex)&(km~s$^{-1}$)&(dex)& (dex) \\
\hline
\endhead
\endfoot
\endlastfoot
\hline
SK-67D05	&-30, 50	& {13.59$^{+0.06}_{-0.08}$}	&-30, 100	&14.19$^{+0.03}_{-0.03}$	&13.43$^{+0.18}_{-0.12}$ 	\\
SK-68D03	&-20, 50	&13.74$^{+0.07}_{-0.06}$	&-20, 100	&14.03$^{+0.01}_{-0.01}$	&12.81$^{+0.18}_{-0.13}$	\\
BI13		&-35, 50	&14.02$^{+0.07}_{-0.06}$	&-35, 100	&14.23$^{+0.06}_{-0.07}$	&12.97$^{+0.21}_{-0.08}$	\\
SK-67D18	&-30, 50	&13.92$^{+0.16}_{-0.11}$	&-30, 100	&14.09$^{+0.12}_{-0.17}$	&12.83$^{+0.27}_{-0.02}$	\\
Sk-67D20	&-30, 50	&13.63$^{+0.10}_{-0.07}$	&-30, 100	&14.05$^{+0.07}_{-0.09}$        &13.61$^{+0.22}_{-0.07}$	\\
PGMW-3070	&-50, 50	&14.09$^{+0.16}_{-0.11}$	&-50, 100	&14.22$^{+0.12}_{-0.16}$	&12.51$^{+0.27}_{-0.01}$	\\
LH103102	&-20, 50	&14.18$^{+0.01}_{-0.01}$	&-20, 100	&14.35$^{+0.02}_{-0.02}$ 	&13.26$^{+0.17}_{-0.13}$	\\
LH91486		&-30, 50	&14.31$^{+0.06}_{-0.05}$	&-30, 100	&14.63$^{+0.05}_{-0.06}$	&14.03$^{+0.20}_{-0.09}$	\\
PGMW-3223	&-50, 50	&13.98$^{+0.08}_{-0.06}$	&-50, 100	&14.01$^{+0.06}_{-0.07}$	&10.06$^{+0.21}_{-0.08}$	\\
HV2241		&-35, 50	&14.31$^{+0.05}_{-0.04}$	&-35, 100	&14.52$^{+0.05}_{-0.05}$	&13.53$^{+0.20}_{-0.10}$	\\
HD32402		&-40, 50	&14.17$^{+0.06}_{-0.05}$	&-40, 100	&14.36$^{+0.07}_{-0.08}$	&13.29$^{+0.07}_{-0.23}$	\\
SK-67D32	&-40, 50	&{13.79$^{+0.10}_{-0.08}$}	&-40, 100	&{13.82$^{+0.09}_{-0.11}$}	&{10.10$^{+0.04}_{-0.24}$}	\\
SK-65D21	&-20, 50	&13.67$^{+0.09}_{-0.07}$	&-20, 100	&13.90$^{+0.07}_{-0.09}$	&12.14$^{+0.23}_{-0.06}$	\\
HV2274		&-20, 50	&14.07$^{+0.10}_{-0.08}$	&-20, 100	&14.18$^{+0.10}_{-0.12}$	&12.43$^{+0.24}_{-0.04}$        \\
Sk-66D51	&-30, 50	&13.83$^{+0.03}_{-0.03}$	&-30, 100	&14.37$^{+0.01}_{-0.01}$	&14.02$^{+0.17}_{-0.14}$         \\
NGC1818-D1	&-50, 50	&14.36$^{+0.03}_{-0.03}$	&-50, 100	&14.45$^{+0.03}_{-0.03}$	&12.69$^{+0.18}_{-0.12}$         \\
SK-70D69	&-50, 50	&{13.66$^{+0.02}_{-0.15}$}	&-50, 100	&{13.96$^{+0.13}_{-0.19}$}	&{11.63$^{+0.04}_{-0.28}$}        \\
Sk-67D69	&-20, 50	&{13.94$^{+0.09}_{-0.07}$}	&-20, 100	&{13.97$^{+0.08}_{-0.09}$}	&{10.28$^{+0.06}_{-0.28}$}         \\
MACHO78-6097	&-30, 50	&14.04$^{+0.05}_{-0.05}$	&-30, 100	&14.29$^{+0.05}_{-0.05}$	&13.21$^{+0.20}_{-0.10}$        \\
BI130		&-30, 50	&{14.09$^{+0.05}_{-0.04}$}	&-30, 100	&{14.28$^{+0.08}_{-0.12}$}	&{13.73$^{+0.03}_{-0.25}$}         \\
SK-69D94	&-20, 50	&13.73$^{+0.04}_{-0.03}$	&-20, 100	&14.12$^{+0.06}_{-0.07}$	&13.53$^{+0.21}_{-0.09}$         \\
SNR0519-697	&2, 50		&14.05$^{+0.06}_{-0.05}$	&2, 100		&14.36$^{+0.05}_{-0.06}$	&13.63$^{+0.20}_{-0.09}$         \\
BREY22		&10, 50		&13.85$^{+0.09}_{-0.08}$	&10, 100	&13.94$^{+0.08}_{-0.10}$	&-0-	\\
HD269445	&-30, 50	&13.71$^{+0.06}_{-0.04}$	&-30, 100	&14.36$^{+0.02}_{-0.03}$	&13.93$^{+0.17}_{-0.13}$         \\
SK-69D124	&-35, 50	&13.34$^{+0.09}_{-0.09}$	&-35, 100	&13.85$^{+0.08}_{-0.10}$	&13.24$^{+0.23}_{-0.05}$         \\
SK-67D105	&-40, 50	&14.17$^{+0.05}_{-0.05}$	&-40, 100	&14.39$^{+0.05}_{-0.05}$	&13.20$^{+0.20}_{-0.10}$         \\
SK-67D106	&-20, 50	&13.79$^{+0.23}_{-0.14}$	&-20, 100	&13.81$^{+0.15}_{-0.22}$	&10.05$^{+0.02}_{-0.27}$        \\
SK-67D107	&-40, 50	&14.52$^{+0.03}_{-0.02}$	&-40, 100	&14.62$^{+0.03}_{-0.03}$	&12.94$^{+0.18}_{-0.12}$         \\
HD36521		&-30, 50	&14.06$^{+0.17}_{-0.12}$	&-30, 100	&14.19$^{+0.13}_{-0.19}$	&13.31$^{+0.28}_{-0.04}$         \\
SK-68D82	&-20, 50	&13.97$^{+0.10}_{-0.07}$	&-20, 100	&14.23$^{+0.07}_{-0.09}$	&13.46$^{+0.23}_{-0.06}$         \\
BI170		&-35, 50	&{13.94$^{+0.03}_{-0.02}$}	&-35, 100	&{14.19$^{+0.02}_{-0.02}$}	&{13.14$^{+0.13}_{-0.18}$}         \\
SK-67D111	&-30, 50	&14.29$^{+0.05}_{-0.04}$	&-30, 100	&14.43$^{+0.05}_{-0.05}$	&13.20$^{+0.20}_{-0.10}$         \\
HV2543		&-35, 50	&14.10$^{+0.03}_{-0.03}$	&-35, 100	&14.27$^{+0.06}_{-0.08}$	&13.12$^{+0.24}_{-0.07}$         \\
SK-70D91	&-45, 50	&14.11$^{+0.04}_{-0.04}$	&-45, 100	&14.34$^{+0.03}_{-0.04}$	&13.46$^{+0.19}_{-0.11}$         \\
Sk-66D100	&-35, 50	&{14.19$^{+0.05}_{-0.04}$}	&-35, 100	&{14.21$^{+0.04}_{-0.05}$}	&{10.11$^{+0.10}_{-0.19}$}         \\
HDE269599	&-40, 50	&14.08$^{+0.04}_{-0.04}$	&-40, 100	&14.18$^{+0.04}_{-0.05}$	&-0-        \\
SK-65D63	&-40, 50	&13.16$^{+0.41}_{-0.23}$	&-40, 100	&13.61$^{+0.16}_{-0.37}$	&12.90$^{+0.33}_{-0.16}$         \\
HV982		&40, 50		&12.55$^{+0.10}_{-0.08}$	&40, 100	&13.86$^{+0.07}_{-0.08}$	&12.97$^{+0.22}_{-0.06}$         \\
SK-70D97	&20, 50		&12.91$^{+0.16}_{-0.11}$	&20, 100	&13.70$^{+0.09}_{-0.12}$	&12.72$^{+0.25}_{-0.03}$         \\
Sk-67D144	&-30, 50	&14.07$^{+0.07}_{-0.06}$	&-30, 100	&14.14$^{+0.06}_{-0.07}$	&13.25$^{+0.21}_{-0.08}$         \\
BI184		&-30, 50	&14.25$^{+0.04}_{-0.04}$	&-30, 100	&14.48$^{+0.04}_{-0.04}$	&13.25$^{+0.21}_{-0.08}$         \\
NGC2004-B15	&-30, 50	&14.11$^{+0.10}_{-0.08}$	&-30, 100	&14.22$^{+0.09}_{-0.12}$	&9.78$^{+0.24}_{-0.03}$          \\
SK-71D38	&-50, 50	&14.29$^{+0.01}_{-0.01}$	&-50, 100	&14.49$^{+0.04}_{-0.04}$	&13.39$^{+0.21}_{-0.08}$         \\
Sk-71D45	&-30, 50	&{13.81$^{+0.73}_{-0.34}$}	&-30, 100	&{14.01$^{+0.36}_{-0.55}$} 	&{13.71$^{+0.41}_{-0.04}$}         \\
SK-67D166	&-20, 50	&14.05$^{+0.04}_{-0.03}$	&-20, 100	&14.28$^{+0.04}_{-0.04}$	&12.80$^{+0.18}_{-0.12}$         \\
SK-67D168	&-30, 50	&13.97$^{+0.02}_{-0.03}$	&-30, 100	&14.24$^{+0.02}_{-0.03}$	&13.04$^{+0.17}_{-0.13}$         \\
SNR0532-675	&-70, 50	&{14.20$^{+0.04}_{-0.04}$}	&-70, 100	&{14.38$^{+0.04}_{-0.05}$}	&{13.12$^{+0.10}_{-0.20}$}         \\
SK-67D191	&-40, 50	&14.45$^{+0.02}_{-0.02}$	&-40, 100	&14.55$^{+0.01}_{-0.01}$	&11.28$^{+0.16}_{-0.14}$         \\
HV5936		&-30, 50	&14.09$^{+0.05}_{-0.04}$	&-30, 100	&14.34$^{+0.04}_{-0.04}$	&13.59$^{+0.17}_{-0.13}$         \\
Sk-69D191	&-30, 50	&14.06$^{+0.04}_{-0.04}$	&-30, 100	&14.22$^{+0.04}_{-0.04}$	&13.90$^{+0.16}_{-0.14}$         \\
J053441-693139	&-35, 50	&13.94$^{+0.23}_{-0.15}$	&-35, 100	&14.17$^{+0.15}_{-0.24}$	&12.65$^{+0.30}_{-0.09}$         \\
Sk-67D211	&-40, 50	&{14.07$^{+0.06}_{-0.05}$}	&-40, 100	&{14.12$^{+0.06}_{-0.06}$}	&{11.44$^{+0.09}_{-0.21}$}         \\
BREY64		&10, 50		&13.29$^{+0.38}_{-0.20}$	&10, 100	&14.21$^{+0.16}_{-0.25}$	&13.12$^{+0.31}_{-0.10}$         \\
SNR0536-692	&38, 50		&13.28$^{+0.39}_{-0.20}$	&38, 100	&14.06$^{+0.49}_{-0.23}$	&13.46$^{+0.38}_{-0.37}$         \\
SK-69D220	&-40, 50	&14.00$^{+0.02}_{-0.02}$	&-40, 100	&14.28$^{+0.02}_{-0.02}$	&13.44$^{+0.17}_{-0.13}$         \\
Sk-66D172	&-30, 50	&13.87$^{+0.04}_{-0.04}$	&-30, 100	&14.32$^{+0.03}_{-0.03}$	&13.95$^{+0.18}_{-0.12}$         \\
BI253		&-20, 50	&13.99$^{+0.05}_{-0.05}$	&-20, 100	&14.34$^{+0.04}_{-0.04}$	&13.02$^{+0.19}_{-0.11}$         \\
SK-68D137	&-20, 50	&13.50$^{+0.04}_{-0.04}$	&-20, 100	&14.33$^{+0.05}_{-0.06}$	&13.36$^{+0.20}_{-0.09}$         \\
MK42		&-20, 50	&13.79$^{+0.12}_{-0.10}$	&-20, 100	&14.40$^{+0.09}_{-0.11}$	&13.71$^{+0.24}_{-0.04}$         \\
SK-69D243	&10, 50		&13.45$^{+0.14}_{-0.11}$	&10, 100	&14.32$^{+0.07}_{-0.09}$	&13.64$^{+0.23}_{-0.06}$         \\
30DOR-S-R136	&-30, 50	&14.03$^{+0.13}_{-0.10}$	&-30, 100	&14.16$^{+0.10}_{-0.16}$	&10.40$^{+0.26}_{-0.01}$         \\
SK-69D246	&-40, 50	&14.25$^{+0.03}_{-0.03}$	&-40, 100	&14.51$^{+0.03}_{-0.03}$	&13.57$^{+0.18}_{-0.12}$        \\
HDE269927	&-40, 50	&14.10$^{+0.05}_{-0.05}$	&-40, 100	&14.32$^{+0.05}_{-0.06}$	&13.12$^{+0.21}_{-0.09}$         \\
SK-69D257	&-40, 50	&14.21$^{+0.06}_{-0.05}$	&-40, 100	&14.45$^{+0.05}_{-0.05}$	&13.44$^{+0.20}_{-0.10}$         \\
SNR0543-689	&0, 50		&13.95$^{+0.06}_{-0.05}$	&0, 100		&14.20$^{+0.06}_{-0.07}$	&13.21$^{+0.22}_{-0.08}$         \\
D301-1005	&-30, 50	&14.16$^{+0.04}_{-0.04}$	&-30, 100	&14.41$^{+0.04}_{-0.04}$	&13.56$^{+0.19}_{-0.11}$         \\
SK-67D250	&-40, 50	&14.24$^{+0.04}_{-0.04}$	&-40, 100	&14.29$^{+0.05}_{-0.04}$	&11.70$^{+0.19}_{-0.10}$         \\
D301-NW8	&-30, 50	&{13.89$^{+0.01}_{-0.01}$}	&-30, 100	&{14.12$^{+0.07}_{-0.82}$}	&{13.06$^{+0.01}_{-0.20}$}         \\
SK-70D115	&-40, 50	&14.26$^{+0.21}_{-0.14}$	&-40, 100	&14.35$^{+0.07}_{-0.09}$	&12.68$^{+0.23}_{-0.06}$        \\
\hline
\end{longtable}
\medskip
\noindent
{ $\dag$ column densities are calculated in the range from 100 km s$^{-1}$ to the upper limit as given in the column 2 of Table \ref{Tab1}
 for each sightline.}}


\newpage

\begin{table}

\caption{O~{\small VI} column density [log N(O~{\small VI})] value from literature}
\begin{tabular}{lccccc}
\label{Tab4}
Region & No. of stars & Mean & Median & Range & Ref.\\
\hline 
Galactic Halo & 11 & 14.45 & 14.47 & 14.16-14.84& a\\
Galactic Halo & 12 & 14.52 & 14.55 & 14.22-14.67& b\\
Galactic Halo & 22 & 14.17 & 14.25 & 13.65-14.57& c\\
Local ISM     &39 & 13.05 & 13.10 & 12.38-13.60 & d\\
Galactic Disk & 148 & 14.23&  --    & 12.90-14.68 & e\\
Galactic Halo & 139 & 14.11& 14.15 &    13.23-15.03 & f\\
\hline
\end{tabular}
\end{table}
\medskip
\noindent
a-\citep{Savage00}, b-\citep{Howk02a}, c-\citep{Zsargo2003}, d-\citep{Savage06}, e-\citep{Bowen et al.}, f-\citep{Savage2009}




\label{lastpage}
\end{document}